# Girth-12 Quasi-Cyclic LDPC Codes with Consecutive Lengths

Guohua Zhang and Xinmei Wang
State Key Lab. of Integrated Service Networks
Xidian University
Xi'an,China
zhangghcast@163.com

*Abstract*— A method to construct girth-12 (3,L) quasi-cyclic low-density parity-check (QC-LDPC) codes with all lengths larger than a certain given number is proposed, via a given girth-12 code subjected to some constraints. The lengths of these codes can be arbitrary integers of the form LP, provided that P is larger than a tight lower bound determined by the maximal element within the exponent matrix of the given girth-12 code. By applying the method to the case of row-weight six, we obtained a family of girth-12 (3,6) QC-LDPC codes for arbitrary lengths above 2688, which includes 30 member codes with shorter code lengths compared with the shortest girth-12 (3,6) QC-LDPC codes reported by O'Sullivan.

## I. INTRODUCTION

A quasi-cyclic low-density parity-check (QC-LDPC) code is defined by the null space of a parity-check matrix which is based on circulant permutation matrices [1]. Compared with randomly constructed codes, QC-LDPC codes can be encoded in linear time with shift registers and require small memory space to store the code graphs for decoding. Therefore, QC-LDPC codes are favorable for hardware implementation. For iterative sum-product decoding of QC-LDPC codes, a part of performance loss is usually attributed to small values of girth; hence, it is desirable to construct QC-LDPC codes having as large a girth as possible [1]-[4]. It has been shown [1] that the girth of any QC-LDPC code is at most 12. Few methods have been known to guarantee a girth of 12 (girth-12) and Tanner's (3,5) QC-LDPC codes [5] and O'Sullivan's methods [6] are two of such constructions. Although the two well-known methods can construct girth-12 QC-LDPC codes, the code lengths of these QC-LDPC codes are more or less limited. Our work was first motivated by this limitation. Instead of presenting a specific method to guarantee girth-12 for some specific code lengths, we propose a way to construct girth-12 (3,L) QC-LDPC codes with all lengths larger than a certain given number, which are termed as consecutive lengths in this paper for simplicity. It should be noted that these codes are all constructed starting from a given girth-12 code subjected to some constraints and not from scratch.

The rest of the paper is organized as follows. A theorem on girth-12 (3,L) QC-LDPC codes with some constraints is derived in Section II. In Section III, a class of girth-12 (3, 6) QC-LDPC codes with consecutive lengths is constructed and the simulation results are given. Section IV concludes the paper.

## II. CODES FROM A SPECIAL TYPE OF EXPONENT MATRICES

### A. Definitions

The exponent matrix $\mathbf{E}$ of a $(3, L)$ QC-LDPC code of length $N = XL$ can be represented by

$$\mathbf{E} = \begin{bmatrix} 0 & 0 & \cdots & 0 \\ p_{1,0} & p_{1,1} & \cdots & p_{1,L-1} \\ p_{2,1} & p_{2,1} & \cdots & p_{2,L-1} \end{bmatrix} \quad (1)$$

where for $1 \le u \le 2, 0 \le v \le L-1, p_{u,v} \in \{0,1,...,X-1\}$, $p_{u,0} = 0$. The corresponding parity-check matrix $\mathbf{H}_X$ can be represented by

$$\mathbf{H}_X = \begin{bmatrix} \mathbf{I}(0) & \mathbf{I}(0) & \cdots & \mathbf{I}(0) \\ \mathbf{I}(p_{1,0}) & \mathbf{I}(p_{1,1}) & \cdots & \mathbf{I}(p_{1,L-1}) \\ \mathbf{I}(p_{2,0}) & \mathbf{I}(p_{2,1}) & \cdots & \mathbf{I}(p_{2,L-1}) \end{bmatrix} \quad (2)$$

where $\mathbf{I}(p)$ represents the circulant permutation matrix [1] of size $X \times X$ with a one at column-$(r+p) \mod X$ for row-$r$, $0 \le r \le X-1$, and zero elsewhere.

We use the notation $g(\mathbf{H}_X)$ to denote the girth of $\mathbf{H}_X$.

### B. Girth-12 (3,L) QC-LDPC Codes of Consecutive Lengths from One Such Code with Some Constraints

The main result of this paper is given by the following theorem.

**Theorem 1**: For a certain integer $Q$, if $\mathbf{H}_Q$ satisfies the following conditions:

(Cond.1) $g(\mathbf{H}_Q) = 12$;

(Cond.2) $p_{1,v} \le p_{2,v}, v = 0,1,...,L-1$;

(Cond.3) $p_{2,x} - p_{2,y} \ge p_{1,x}$, where $p_{2,x}$ and $p_{2,y}$ are the largest and second largest values, respectively,

This work was supported by the National Basic Research Program (973) of China under Grants 2010CB328300, by the National Natural Science Foundation of China under Grants U0635003, and by the 111 Program of China under Grant B08038.

within $\{p_{2,0}, p_{2,1}, ..., p_{2,L-1}\}$, and $p_{1,x}$ is the largest value within $\{p_{1,0}, p_{1,1}, ..., p_{1,L-1}\}$, then for any arbitrary integer $P \geq 2p_{2,x} + 1$, $g(\mathbf{H}_P) = 12$.

To prove the preceding theorem, we investigate the existence of cycles of lengths 4, 6, 8, and 10 within $\mathbf{H}_P$ (or equivalently within $\mathbf{E}$).

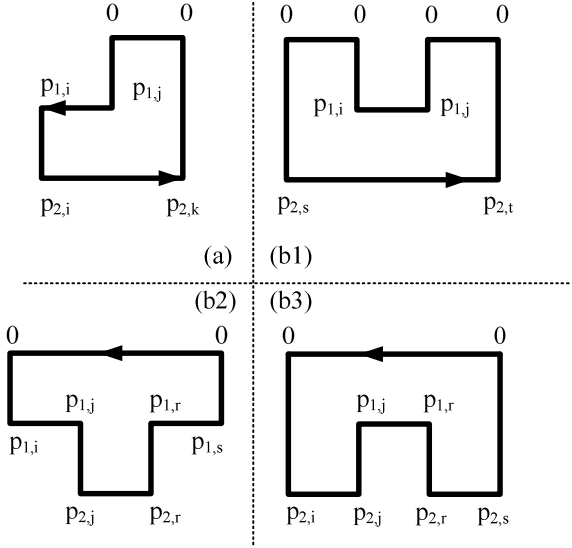

Figure 1.  All possible patterns of 6-cycles and 8-cycles within three rows of $\mathbf{E}$

*1) Case A: cycles of length four*

Since there is no 4-cycle within $\mathbf{H}_Q$, $p_{1,i} \neq p_{1,j}$ for $i \neq j$. Thus, since $P > p_{1,x}$, $p_{1,i} \neq p_{1,j} (\mod P)$, which implies the nonexistence of 4-cycles within the first and second rows of $\mathbf{E}$. Similarly, since $P > p_{2,x}$ there is no 4-cycle within the first and third rows of $\mathbf{E}$.

Assume that there is a 4-cycle within the second and third rows of $\mathbf{E}$. Then, there exist two integers $i \neq j$ such that

$$(p_{1,i} - p_{2,i}) + (p_{2,j} - p_{1,j}) = 0 (\mod P) \quad (3)$$

Since $P > p_{2,x} + p_{1,x}$, Equation (3) can be simplified to

$$p_{1,i} + p_{2,j} = p_{1,j} + p_{2,i} \quad (4)$$

As a result

$$(p_{1,i} - p_{2,i}) + (p_{2,j} - p_{1,j}) = 0 (\mod Q) \quad (5)$$

Equation (5) defines a 4-cycle of $\mathbf{H}_Q$, which contradicts $g(\mathbf{H}_Q) = 12$.

*2) Case B: cycles of length six*

Since 6-cycles cannot exist within any two rows of $\mathbf{E}$, assume that there is a 6-cycle within the first, second and third rows of $\mathbf{E}$, as shown in Fig.1 (a). Then, there exist three integers $i, j, k (i \neq j; j \neq k; k \neq i)$ such that

$$(0 - p_{1,j}) + (p_{1,i} - p_{2,i}) + (p_{2,k} - 0) = 0 (\mod P) \quad (6)$$

Since $P > p_{2,x} + p_{1,x}$, Equation (6) can be simplified to

$$p_{1,i} + p_{2,k} = p_{1,j} + p_{2,i} \quad (7)$$

Thus,

$$(0 - p_{1,j}) + (p_{1,i} - p_{2,i}) + (p_{2,k} - 0) = 0 (\mod Q) \quad (8)$$

Equation (8) defines a 6-cycle of $\mathbf{H}_Q$, which contradicts $g(\mathbf{H}_Q) = 12$.

*3) Case C: cycles of length eight*

An 8-cycle can exist in any two rows of $\mathbf{E}$ or in three rows of $\mathbf{E}$.

(C.1) Assume that there is an 8-cycle within the first and third rows of $\mathbf{E}$. Then there exist four integers $i, j, k, l (i \neq j; j \neq k; k \neq l; l \neq i)$ such that

$$(0 - p_{2,i}) + (p_{2,j} - 0) + (0 - p_{2,k}) + (p_{2,l} - 0)$$
$$= 0 (\mod P) \quad (9)$$

Since $P > 2p_{2,x}$, Equation (9) can be simplified to

$$p_{2,j} + p_{2,l} = p_{2,i} + p_{2,k} \quad (10)$$

As a result,

$$(0 - p_{2,i}) + (p_{2,j} - 0) + (0 - p_{2,k}) + (p_{2,l} - 0)$$
$$= 0 (\mod Q) \quad (11)$$

Equation (11) defines an 8-cycle of $\mathbf{H}_Q$, which contradicts $g(\mathbf{H}_Q) = 12$.

(C.2) Similarly, since $P > 2p_{2,1}$ there is no 8-cycle within the first and second rows of $\mathbf{E}$.

(C.3) Assume that there is an 8-cycle within the second and third rows of $\mathbf{E}$. Then there exist four integers $i, j, k, l (i \neq j; j \neq k; k \neq l; l \neq i)$ such that

$$(p_{1,i} - p_{2,i}) + (p_{2,j} - p_{1,j}) + (p_{1,k} - p_{2,k}) + (p_{2,l} - p_{1,l})$$
$$= 0 (\mod P) \quad (12)$$

By (Cond.2) and $P > 2p_{2,x}$, Equation (12) can be simplified to

$$(p_{2,j} - p_{1,j}) + (p_{2,l} - p_{1,l}) = (p_{2,i} - p_{1,i})$$
$$+ (p_{2,k} - p_{1,k}) \quad (13)$$

Thus,

$$(p_{1,i} - p_{2,i}) + (p_{2,j} - p_{1,j}) + (p_{1,k} - p_{2,k})$$
$$+ (p_{2,l} - p_{1,l}) = 0 \pmod{Q} \quad (14)$$

Equation (14) defines an 8-cycle of $\mathbf{H}_Q$, which contradicts $g(\mathbf{H}_Q) = 12$.

(C.4) Assume there is an 8-cycle within the first, second and third rows of $\mathbf{E}$. Then the 8-cycle has one of the three distinct patterns shown in Fig.1 (b1-b3).

(C.4.1) If the 8-cycle appears in pattern (b1), then there exist four integers $i, s, t, j (i \neq s; s \neq t; t \neq j; j \neq i)$ such that
$$(p_{1,i} - 0) + (0 - p_{2,s}) + (p_{2,t} - 0) + (0 - p_{1,j})$$
$$= 0 \pmod{P} \quad (15)$$

Since $P > p_{2,x} + p_{1,x}$, Equation (15) can be simplified to $p_{1,i} + p_{2,t} = p_{2,s} + p_{1,j}$ so that
$$(p_{1,i} - 0) + (0 - p_{2,s}) + (p_{2,t} - 0) + (0 - p_{1,j})$$
$$= 0 \pmod{Q} \quad (16)$$

Equation (16) defines an 8-cycle of $\mathbf{H}_Q$, which contradicts $g(\mathbf{H}_Q) = 12$.

(C.4.2) If the 8-cycle appears in pattern (b2), then there exist four integers $i, j, r, s (i \neq j; j \neq r; r \neq s; s \neq i)$ such that
$$(0 - p_{1,i}) + (p_{1,j} - p_{2,j}) + (p_{2,r} - p_{1,r})$$
$$+ (p_{1,s} - 0) = 0 \pmod{P} \quad (17)$$

By (Cond.2) and $P > p_{2,x} + p_{1,x}$, Equation (17) can be simplified to
$$(p_{2,r} - p_{1,r}) + p_{1,s} = (p_{2,j} - p_{1,j}) + p_{1,i} \quad (18)$$
As a result,
$$(0 - p_{1,i}) + (p_{1,j} - p_{2,j}) + (p_{2,r} - p_{1,r})$$
$$+ (p_{1,s} - 0) = 0 \pmod{Q} \quad (19)$$

Equation (19) defines an 8-cycle of $\mathbf{H}_Q$, which contradicts $g(\mathbf{H}_Q) = 12$.

(C.4.3) If the 8-cycle appears in pattern (b2), then there exist four integers $i, j, r, s (i \neq j; j \neq r; r \neq s; s \neq i)$ such that
$$(0 - p_{2,i}) + (p_{2,j} - p_{1,j}) + (p_{1,r} - p_{2,r})$$
$$+ (p_{2,s} - 0) = 0 \pmod{P} \quad (20)$$

By (Cond.2) and $P > 2p_{2,x}$, Equation (20) can be simplified to
$$(p_{2,j} - p_{1,j}) + p_{2,s} = (p_{2,r} - p_{1,r}) + p_{2,i} \quad (21)$$
Consequently,

$$(0 - p_{1,i}) + (p_{1,j} - p_{2,j}) + (p_{2,r} - p_{1,r})$$
$$+ (p_{1,s} - 0) = 0 \pmod{Q} \quad (22)$$

Equation (22) defines an 8-cycle of $\mathbf{H}_Q$, which contradicts $g(\mathbf{H}_Q) = 12$.

*4) Case D: cycles of length ten*

Since 10-cycles cannot exist within any two rows of $\mathbf{E}$, assume that there is a 10-cycle within the first, second and third rows of $\mathbf{E}$. Then the 10-cycle has one of the three distinct patterns depicted in Fig.2.

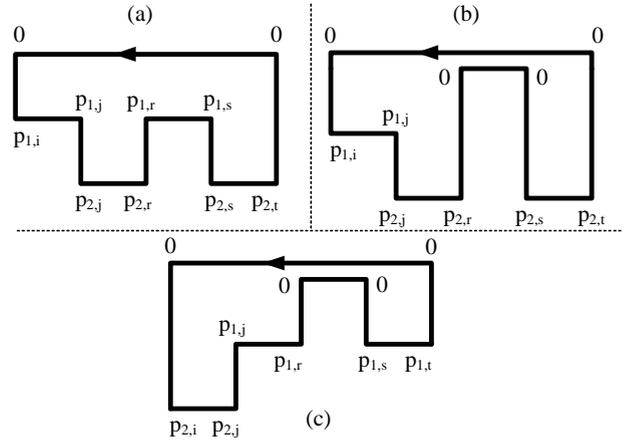

Figure 2. All possible patterns of 10-cycles within $\mathbf{E}$

(D.1) If the 10-cycle appears in pattern (a), then there exist five integers $i, j, r, s, t (i \neq j; j \neq r; r \neq s; s \neq t; t \neq i)$ such that
$$(0 - p_{1,i}) + (p_{1,j} - p_{2,j}) + (p_{2,r} - p_{1,r}) + (p_{1,s} - p_{2,s})$$
$$+ (p_{2,t} - 0) = 0 \pmod{P} \quad (23)$$
Equation (23) can be expressed as
$$(p_{2,r} - p_{1,r}) + p_{2,t} = p_{1,i} + (p_{2,j} - p_{1,j})$$
$$+ (p_{2,s} - p_{1,s}) \pmod{P} \quad (24)$$

Define *RHS* as the right-hand side of equation (24).

(D.1.1) if $s \neq x$ and $j \neq x$:
$$P - RHS > -p_{1,i} + p_{1,x} + p_{1,j} + p_{1,x} + p_{1,s} > 0$$
(D.1.2) if $s = x$ and $j \neq x$:
$$P - RHS > -p_{1,i} + p_{1,x} + p_{1,j} + 0 + p_{1,x} > 0$$
(D.1.3) if $s \neq x$ and $j = x$:
$$P - RHS > -p_{1,i} + 0 + p_{1,x} + p_{1,x} + p_{1,s} > 0$$
(D.1.4) if $s = x$ and $j = x$:
$$P - RHS > -p_{1,i} + 0 + p_{1,x} + 0 + p_{1,x} > 0$$

Since $P > RHS$, equation (24) can be simplified to
$$(p_{2,r} - p_{1,r}) + p_{2,t} = p_{1,i} + (p_{2,j} - p_{1,j}) + (p_{2,s} - p_{1,s}) \quad (25)$$
As a result,
$$(0 - p_{1,i}) + (p_{1,j} - p_{2,j}) + (p_{2,r} - p_{1,r}) + (p_{1,s} - p_{2,s}) + (p_{2,t} - 0) = 0 (\mod Q) \quad (26)$$

Equation (26) defines a 10-cycle of $\mathbf{H}_Q$, which contradicts $g(\mathbf{H}_Q) = 12$.

(D.2) If the 10-cycle appears in pattern (b), then there exist five integers $i, j, r, s, t (i \neq j; j \neq r; r \neq s; s \neq t; t \neq i)$ such that
$$(0 - p_{1,i}) + (p_{1,j} - p_{2,j}) + (p_{2,r} - 0) + (0 - p_{2,s}) + (p_{2,t} - 0) = 0 (\mod P) \quad (27)$$

Equation (27) can be expressed as
$$p_{2,r} + p_{2,t} = p_{2,s} + p_{1,i} + (p_{2,j} - p_{1,j})(\mod P) \quad (28)$$

Define $RHS$ as the right-hand side of equation (28).

(D.2.1) if $s \neq x$ and $j \neq x$:
$$P - RHS > p_{1,x} - p_{1,i} + p_{1,x} + p_{1,j} > 0$$
(D.2.2) if $s = x$ and $j \neq x$:
$$P - RHS > 0 - p_{1,i} + p_{1,x} + p_{1,j} \geq 0$$
(D.2.3) if $s \neq x$ and $j = x$:
$$P - RHS > p_{1,x} - p_{1,i} + 0 + p_{1,x} > 0$$
(D.2.4) if $s = x$ and $j = x$:
$$P - RHS > 0 - p_{1,i} + 0 + p_{1,x} > 0$$
Since $P > RHS$, equation (28) can be simplified to
$$p_{2,r} + p_{2,t} = p_{2,s} + p_{1,i} + (p_{2,j} - p_{1,j}) \quad (29)$$
Consequently,
$$p_{2,r} + p_{2,t} = p_{2,s} + p_{1,i} + (p_{2,j} - p_{1,j})(\mod Q) \quad (30)$$

Equation (30) defines a 10-cycle of $\mathbf{H}_Q$, which contradicts $g(\mathbf{H}_Q) = 12$.

(D.3) If the 10-cycle appears in pattern (c), then there exist five integers $i, j, r, s, t (i \neq j; j \neq r; r \neq s; s \neq t; t \neq i)$ such that
$$(0 - p_{2,i}) + (p_{2,j} - p_{1,j}) + (p_{1,r} - 0) + (0 - p_{1,s}) + (p_{1,t} - 0) = 0 (\mod P) \quad (31)$$

Equation (31) can be expressed as
$$p_{1,s} + p_{2,i} = (p_{2,j} - p_{1,j}) + p_{1,r} + p_{1,t}(\mod P) \quad (32)$$

Define $RHS$ as the right-hand side of equation (32).

(D.3.1) if $j \neq x$:
$$P - RHS > p_{1,x} + p_{1,x} + p_{1,j} - p_{1,r} - p_{1,t} > 0$$
(D.3.2) if $j = x$:
$$P - RHS > p_{2,x} + p_{1,x} - p_{1,r} - p_{1,t} > 0$$
Since $P > RHS$, equation (32) can be simplified to
$$p_{1,s} + p_{2,i} = (p_{2,j} - p_{1,j}) + p_{1,r} + p_{1,t} \quad (33)$$
Thus,
$$(0 - p_{2,i}) + (p_{2,j} - p_{1,j}) + (p_{1,r} - 0) + (0 - p_{1,s}) + (p_{1,t} - 0) = 0(\mod Q) \quad (34)$$

Equation (34) defines a 10-cycle of $\mathbf{H}_Q$, which contradicts $g(\mathbf{H}_Q) = 12$.

It follows from the results of cases A to D that for $P \geq 2p_{2,x} + 1$, $g(\mathbf{H}_P) = 12$, which completes the proof.
□

Let $i = k = 0$ and $j = l = x$. Then equation (9) becomes
$$(0 - 0) + (p_{2,x} - 0) + (0 - 0) + (p_{2,x} - 0) = 0(\mod P) \quad (35)$$

Equation (35) defines an 8-cycle of $\mathbf{H}_P$ for $P = 2p_{2,x}$. This suggests that $2p_{2,x}$ is a tight lower bound above which $\mathbf{H}_P$ always achieves the girth 12.

## III. AN EXAMPLE OF GIRTH-12 QC-LDPC CODES WITH CONSECUTIVE LENGTHS

In order to illustrate the applications of Theorem 1, we give an example for girth-12 QC-LDPC codes with consecutive lengths. By combining a greedy search method and a simulated annealing algorithm (which is not the issue of this paper), we found an exponent matrix $\mathbf{E}$ as follows.

$$\mathbf{E} = \begin{bmatrix} 0 & 0 & 0 & 0 & 0 & 0 \\ 0 & 3 & 14 & 18 & 24 & 26 \\ 0 & 19 & 62 & 107 & 170 & 224 \end{bmatrix} \quad (36)$$

It is easily verified that for $P = 393$ the exponent matrix corresponds to a (3,6) QC-LDPC code with girth 12 and length $N = 2358$. Based on Theorem 1, the exponent matrix $\mathbf{E}$ can be used to construct girth-12 (3,6) QC-LDPC codes with arbitrary lengths $N = 6P$ ($P > 448$). This is an encouraging result, considering that, the up-to-date shortest girth-12 (3,6) QC-LDPC code, to the best of our knowledge, is reported with length 2874 by O'Sullivan [6], while our new family of (3,6) QC-LDPC codes have consecutive lengths above the lower bound of 2688.

In what follows, we show that the girth-12 (3,6) QC-LDPC codes based on the above exponent matrix $\mathbf{E}$ often have no serious performance degradation, as compared with the LDPC codes constructed by progressive-edge-growth (PEG) algorithm [7].In Fig. 3 and 4, we plot the error performance for two girth-12 (3,6) QC-LDPC codes (*P*=449 and 500,respectively) and their counterparts with iterative sum-product decoding. Both QC- and PEG-LDPC codes are of the same rate 0.5. The number of frames in simulation is large enough to ensure that at least 50 error frames appeared at each simulated SNR point. The maximum number of iterations is set to 80.It is shown that compared with random PEG-LDPC codes, the first QC-LDPC code has a little performance degradation (less than 0.1dB at BER=1e-5), while the second has no degradation. The slight performance degradation of the QC-LDPC codes should be well compensated by their good algebraic structures, which are attractive for commercial use and implementation purpose.

Finally, we give some remarks on the minimum distance ($d_{min}$) of the above girth-12 (3,6) QC-LDPC codes. For girth-8 (3,L) QC-LDPC codes, only $d_{min} \geq 6$ is guaranteed and for girth-12 (3,L) QC-LDPC codes, $d_{min}$ at least 14 is guaranteed [7].Using the improved impulse method of [8], we found $d_{min} \leq 24$ for both the above code with P=393 and the above code with P=500, which is the maximum possible $d_{min}$ for (3,L) QC-LDPC codes. This implies that the codes constructed above have significant values in the sense of minimum distance. It would be interesting to find out whether the extension (with difference values of P) also preserves $d_{min}$, or conditions to do so. This problem deserves to be further investigated and we leave it as an open problem.

## IV. CONCLUSION

This paper presents a way to generate girth-12 (3, L) QC-LDPC codes for all lengths larger than a certain given number from one such code subjected to some constrains. The main advantage of Theorem 1 is that it avoids the repetitions of executing a construction procedure (often combined with extensive computer search) for different values of P if only a girth of 12 is pursued. Finally, we note that Theorem 1 can be naturally applied to the construction of girth-12 column-weight-three codes with essentially arbitrary lengths, which have guaranteed error-correction capability and can correct any combinations of at most five errors in six iterations using Gallager-A algorithm, according to the main results of an unpublished literature [9].

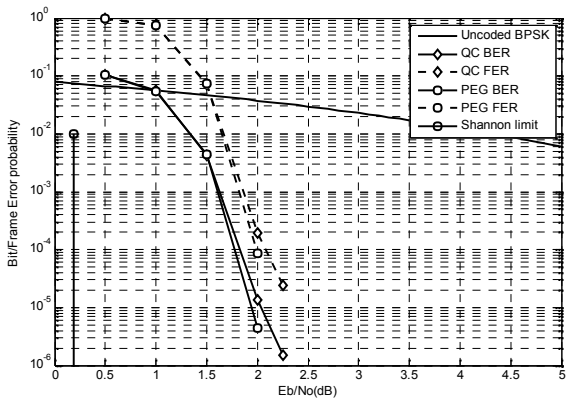

Figure 3. Error performance of the (2694, 1349) QC-LDPC code and the (2694,1347) PEG-LDPC code

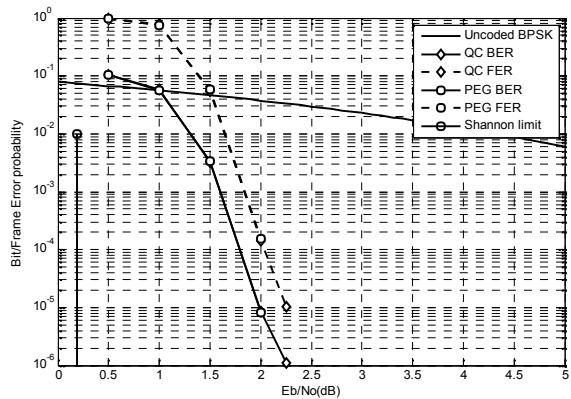

Figure 4. Error performance of the (3000, 1502) QC-LDPC code and the (3000,1500) PEG-LDPC code